# Inversionless gain in a lossy medium

Eliran Talker, Yefim Barash, Noa Mazurski and Uriel Levy*

Department of Applied Physics, The Faculty of Science, The Center for Nanoscience and Nanotechnology, The Hebrew University of Jerusalem, Jerusalem 91904, Israel

*Corresponding author: ulevy@mail.huji.ac.il

## Abstract

We study gain without inversion due to coherence effects in a Doppler-broadened degenerate three-level system of a rubidium-hydrogen mixture in a miniaturized micron scale custom vapor cell. The cell miniaturization gives rise to collisions of atoms with the walls of the cell. This, combined with the high collision rate with the hydrogen buffer gas allows us to observe gain in the absorption spectra. Furthermore, we analyze the role of cell miniaturization in the evolution of the gain profile. In addition to fundamental interest, the observation of gain without inversion in our miniaturized cells paves the way for applications such as miniaturized lasers without inversion.

## Introduction

Atomic coherence effects such as a coherent population trapping (CPT) [1–7], electromagnetically induced transparency (EIT) [5,8–15], high index of refraction with vanishing absorption, and amplification and lasing without inversion (AWI and LWI) have been intensively investigated over recent years due to their fundamental properties and their multiple potential applications [16–24], ranging from laser cooling to isotope separation and from ultrahigh-sensitive magnetometers to the generation of giant pulses of laser light. Among these phenomena, AWI and LWI have received considerable attention for their potential to obtain laser light in diverse spectral domains, e.g., the X-ray range where conventional methods based on population inversion are not available or are difficult to implement.

In the present work, we demonstrate for the first-time gain without population inversion in a micron-scale Rubidium vapor cell without the use of an incoherent pump beam. The demonstrated effects take advantage of the frequent atom-wall collisions in our micrometer cell as well as the frequent collisions with the buffer gas (Hydrogen) to suppress the optical pumping. Below we describe the theoretical background and the simulations of gain and populations in different types of cells. This is followed by describing the experimental setup and the obtained experimental results.

# Theory and simulations

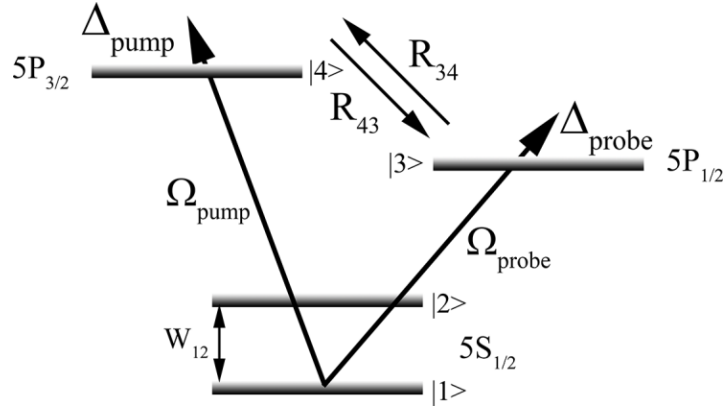

FIG.1. Schematic diagram level of Rubidium vapor, $W_{12}$ is the population exchange rate of the ground levels $\Omega_{probe}, \Omega_{pump}$ is the Rabi frequency of the probe and the pump beam respectively, $R_{34}, R_{43}$ is the energy transfers rates between excited states due to buffer gas collisions, $\Delta_{probe}, \Delta_{pump}$ is the probe and pump detuning.

To analyze of Rubidium (Rb) hot vapor system, in the presence of buffer gas and wall collisions, we consider a model as shown in FIG 1 which consists of four level system labeled $|1\rangle, |2\rangle, |3\rangle$ and $|4\rangle$. A strong driving field induces a coherent coupling between levels $|1\rangle$ and $|4\rangle$, whereas a weak probe beam is scanned around the transition between levels $|1\rangle$ and $|3\rangle$. Under the rotating wave approximation, the Hamiltonian of the system, including the interaction between the atoms and the two electromagnetic fields and can be written as:

$$\mathcal{H} = \mathcal{H}_1 + \mathcal{H}_2$$

(1)
$$\mathcal{H}_1 = \hbar\omega_{pr} a_3^+ a_3 + \hbar\omega_{pu} a_4^+ a_4$$

$$\mathcal{H}_2 = \hbar\Delta_{pr} a_3^+ a_3 + \hbar\Delta_{pu} a_4^+ a_4 + \hbar\left[\Omega_{pr} a_3^+ a_1 + \Omega_{pu} a_4^+ a_1 + c.c.\right]$$

Where $\Delta_{pr} = \omega_{pr} - \omega_{13}$, $\Delta_{pu} = \omega_{pu} - \omega_{14}$, are the detunings whereas $\omega_{13}$ and $\omega_{14}$ denote the transition frequencies from $|1\rangle$ to $|3\rangle$ and $|1\rangle$ to $|4\rangle$, respectively. $\omega_{pr}, \omega_{pu}$ are the probe and the pump beam frequencies respectively. $\Omega_{pr}, \Omega_{pu}$ are the Rabi frequencies of the probe and the pump fields. The time evolution of the density operators is described by the Liouville equation:

(2)
$$\partial_t \rho = -\frac{i}{\hbar}[\mathcal{H}, \rho] - \frac{1}{2}\{\Gamma, \rho\} + \Lambda$$

Where the square brackets denote the commutator, and the curly brackets stand for the anticommutator. We assume that the upper states $|3\rangle$ and $|4\rangle$ decays spontaneously with rates $\gamma_3$ and $\gamma_4$ respectively. In addition to the spontaneous relaxation, the excited states transfer population due to collisional transfer, as well as by pressure broadening relaxation, both induced by the buffer gas (note $R_{34}$ and $R_{43}$). The overall relaxation is described by the matrix $\Gamma$:

(3) $$\Gamma = W_{12}|1\rangle\langle 1| + W_{12}|2\rangle\langle 2| + (\gamma_3 + R_{34})|3\rangle\langle 3| + (\gamma_3 + R_{43})|4\rangle\langle 4|$$

$W_{12}$ is the wall collision relaxation (see Appendix). $\gamma_3$ and $\gamma_4$ are the $5P_{1/2}$ and $5P_{3/2}$ states relaxation rate respectively. $R_{34}, R_{43}$ are the collisional transfer rates between $5P_{1/2}$ and $5P_{3/2}$ levels due to the buffer gas (for more information see Appendix). We use here the simple model of ground state relaxation. $\Lambda$ is related to the repopulated matrix (see Appendix). The evolution equation for the atomic coherence related to the absorption as obtained from equation (2) is equal to (for more details see Appendix)

(4) $$\rho_{13} = \frac{2\Omega_{pr} i(\rho_{33}-\rho_{11})+2i\Omega_{pu}\rho_{43}}{R+W_{12}+\gamma_3+2i\Delta_{pr}}$$

As derived from equation (4) the gain is dependent on the coherence of the excited states ($\rho_{43}$) and the population difference between the excited state $|3\rangle$ and the population of the ground state $|1\rangle$ ($\rho_{33} - \rho_{11}$).

Fig. 2(a). Shows the evolution of the ground states and the excited states population as a function of time. The relaxation rate ($W_{12}$) between the ground states due to wall collisions was assumed to be $W_{12} \sim 0.5 \cdot \gamma_3$, corresponding to ~8 Torr of $H_2$ buffer gas. The wall collision suppresses the optical pumping. Therefore, the population difference $\rho_{33} - \rho_{11}$ is determined by the rates which atoms leave and decay into states $|1\rangle$ and $|2\rangle$. In the limit of a weak probe beam, atoms can leave state $|1\rangle$ only by being pumped to state $|4\rangle$. It is worth mentioning that if level $|2\rangle$ decays faster than the level $|4\rangle$, an inversion of the transition coupled by the weak probe field cannot be achieved (see Fig. 2(b)). By observing the steady states values, we see that the value of the $\rho_{13}$ is positive (Fig. 2(b ,c)), hence gain is achieved in spite of the fact that $\rho_{33} - \rho_{11}$ is negative, giving rise to gain without inversion.

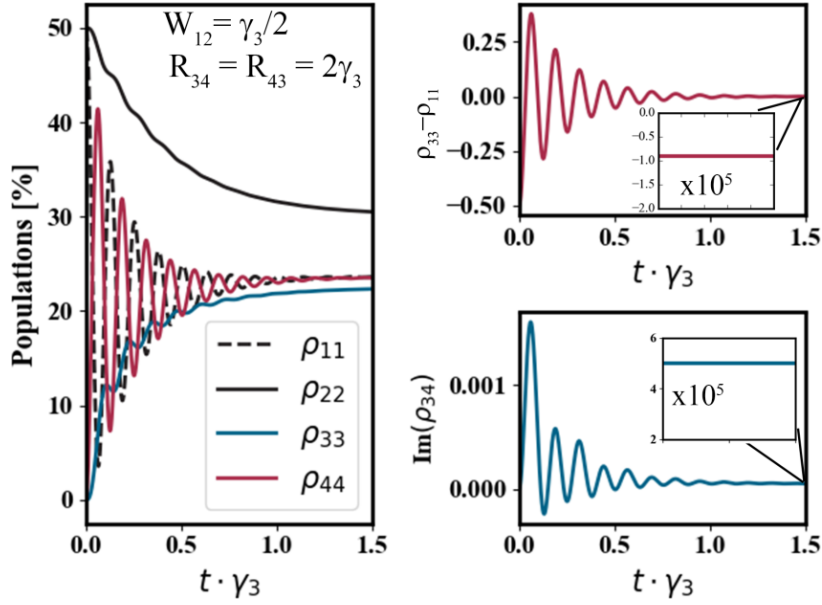

FIG. 2. Time evolution of the density matrix as calculated from the four level optical Bloch model, with parameters of $\Omega_{probe} = 0.05 \cdot \gamma_3$ and $\Omega_{pump} = 60 \cdot \gamma_3$, $W_{12} = \gamma_3/2$ and $R_{34} = R_{43} = 2 \cdot \gamma_3$ and $\gamma_3 = 2\pi \cdot 5.75\ MHz$

We now address the case of a large cell with negligible wall collisions by repeat the calculations, this time with $W_{12} = 0$. Fig. 3(a) shows that when the wall collision is negligible the population of the ground state $|2\rangle$ is increased due to optical pumping. As a result, the coherence of the levels $|3\rangle$ and $|4\rangle$ is diminished (i.e. $\rho_{43}$ is nulled, see Fig. 3(c)), and $\rho_{13}$ become negative. As such, observing gain without inversion becomes impossible in this case.

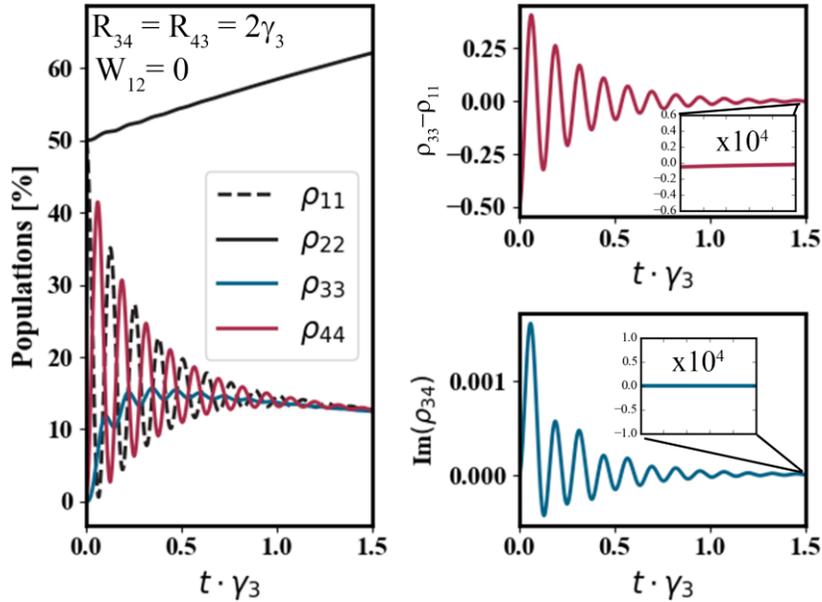

FIG. 3. Time evolution of the density matrix as calculated from the four level optical Bloch model, with parameters of $\Omega_{probe} = 0.05 \cdot \gamma_3$ and $\Omega_{pump} = 60 \cdot \gamma_3$, $W_{12} = 0$ and $R_{34} = R_{43} = 2 \cdot \gamma_3$ and $\gamma_3 = 2\pi \cdot 5.75\ MHz$

Next, we calculate the gain (absorption) coefficient as a function of probe detuning for a weak probe field on the $|1\rangle$ to $|3\rangle$ transition. As mentioned before, the gain is proportional to the imaginary part of the off-diagonal elements of the atomic density matrix $\rho_{13}$. In the weak probe field, and under condition of pump at resonance, we obtain the following expression for the frequency dependent susceptibility:

$$(5) \quad \chi(\Delta_{pr}, v) = \frac{3\lambda_{pr}^2 NL\gamma_3}{4\pi} \frac{Im(\rho_{13}(\Delta_{pr}, v))}{\Omega_{pr}} =$$

$$\frac{3\lambda_{pr}^2 NL\gamma_3}{4\pi\Omega_{pr}} \frac{2(R + W_{12} + \gamma_3)}{(R + W_{12} + \gamma_3)^2 + (2\Delta_{pr})^2} \left( \Omega_{pu}\rho_{43}(\Delta_{pr}, v) + \Omega_{pr}(\rho_{33}(\Delta_{pr}, v) - \rho_{11}(\Delta_{pr}, v)) \right)$$

From equation (5) (see right parenthesis of the second equality) we can identify two contributions which affect the gain\absorption terms. The first is related to the coherence of the two excited states, $\rho_{43}$, while the second corresponds to the population inversion of the ground state and the excited state of the probe beam, $\rho_{33} - \rho_{11}$. Therefore, as mentioned before, we can observe gain in the absorption although there is no population inversion (i.e. $\rho_{33} - \rho_{11} < 0$). In order to calculate the actual gain taking into account the Doppler effect which is a dominant factor in our hot vapor system, we multiply the susceptibility by the Boltzmann distribution and integrating over all velocities

$$(6) \quad G(\Delta_{pr}) = \int_{-\infty}^{\infty} \chi(\Delta_{pr}, v) \cdot \exp\left(-\frac{v^2}{u^2}\right) dv$$

Where u is the most probable velocity.

Fig. 4 shows transmission spectra for the two different cases of negligible collisions ($W_{12} = 0$, red line), and significant collisions ($W_{12} = 0.5 \cdot \gamma_3$, blue line). Indeed, when the collision effect is negligible, we cannot observe gain in the transmission, whereas when the wall collision rate is significant one can observe gain within the transmission spectra.

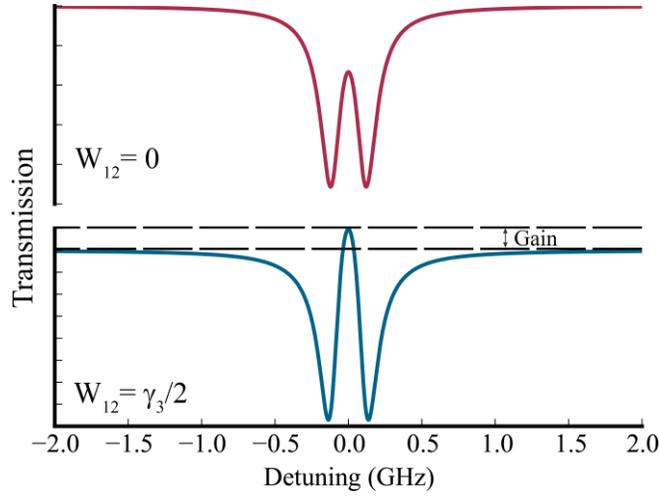

FIG.4. Simulation of transmission spectra as a function of the probe detuning with and without ground state relaxation ($W_{12}$) due to wall collision. In both of the plots the parameter are $\Omega_{probe} = 0.05 \cdot \gamma_3$ and $\Omega_{pump} = 60 \cdot \gamma_3$, and $R_{34} = R_{43} = 2 \cdot \gamma_3$ and $\gamma_3 = 2\pi \cdot 5.75\ MHz$

# Experimental Setup and results

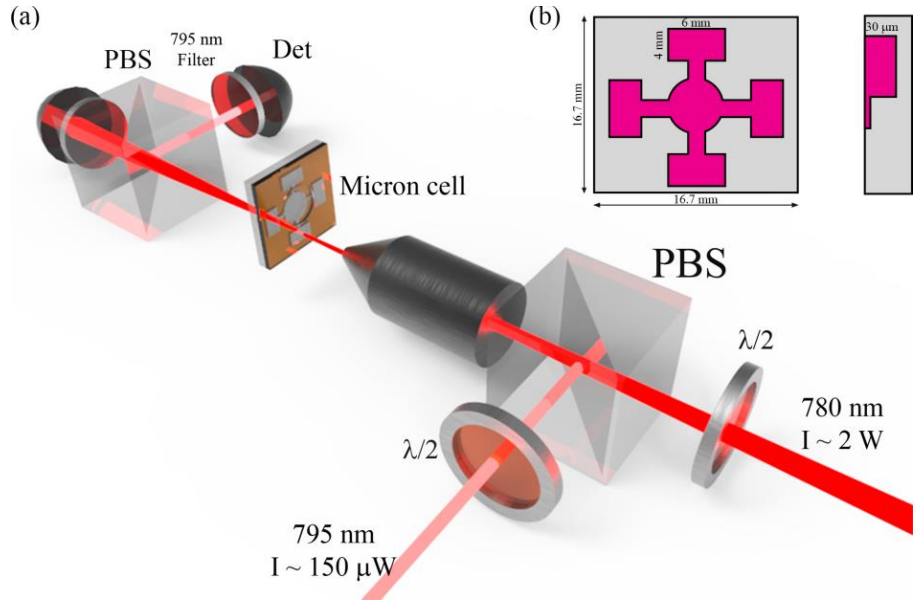

FIG. 5. (a) cartoon of the experimental setup used for the entire results. PBS - polarizing beam splitter, Det – detector (b) top view of the device and side cross-section schematic (not to scale) of the micrometer rubidium cell.

An illustration of our experimental setup is shown in Fig. 5(a), consisting of two laser beams. The pump beam wavelength is 780 nm, corresponding to the D$_2$ transition ($5S_{1/2} - 5P_{3/2}$) of the Rb vapor. This laser has a short-term linewidth on the order of 100 kHz, and its frequency is stabilized to the $F_g = 2 \rightarrow F_e = 3$ transition using the well-known saturated absorption spectroscopy (SAS) technique [25]. As shown in the previous section, to obtain gain we need to generate a strong excited state coherence. This is done by setting the pump intensity to be fairly high, ~2 W, by using a tapered amplifier (Toptica BOOSTA). The probe beam is at tuned to the wavelength of 795 nm corresponding to the $D_1$ transition ($5S_{1/2} - 5P_{1/2}$) of Rb. It was kept at a relatively low intensity level around 150 µW. At such intensity levels we don't change the population distribution of the system, but only measure the probe transmission though the miniaturized cell, under the presence of wall collisions, and buffer gas and the at conditions of strong pump beam intensity. The linewidth of the probe beam laser is ~100 kHz.

As mentioned before, in order to observe gain we need the wall collision rate to be significant with respect to the excited state decay rate. To facilitate such a case, we have fabricated a custom miniaturized vapor cell with thickness of about 30 microns along the light propagation direction. Schematic description of the cell geometry is given in FIG. 5b. For more details related to the cell fabrication method see [26].

To eliminate the Zeeman sub levels splitting, the residual magnetic field was canceled using fourfold Mumetal shield which reduced the residual magnetic field to less than $10^{-7}$ T. The experiments were done at elevated temperatures in the range of 150°C-250°C corresponding to atomic density of $3.5 \cdot 10^{19}\ [m^{-3}]$ and $3.05 \cdot 10^{20}\ [m^{-3}]$ respectively. These high temperatures allow us to reach high optical densities even at low propagation length. Furthermore, by increasing the temperature we increase the most probable velocity and as a result the wall collision rate is increased, further enhancing the effect of gain without inversion.

In order to detect only the probe beam while eliminating the detection of the pump beam, we set the pump beam to be TE polarized while the probe beam was set to be TM polarized. By using a polarizing beam splitter (PBS) after the vapor cell we filtered out the TE mode and detected only the TM mode. To further improve the filtering, we have added a bandpass filter around 795 nm allowing only the probe to be transmitted, while rejecting the pump at 780 nm.

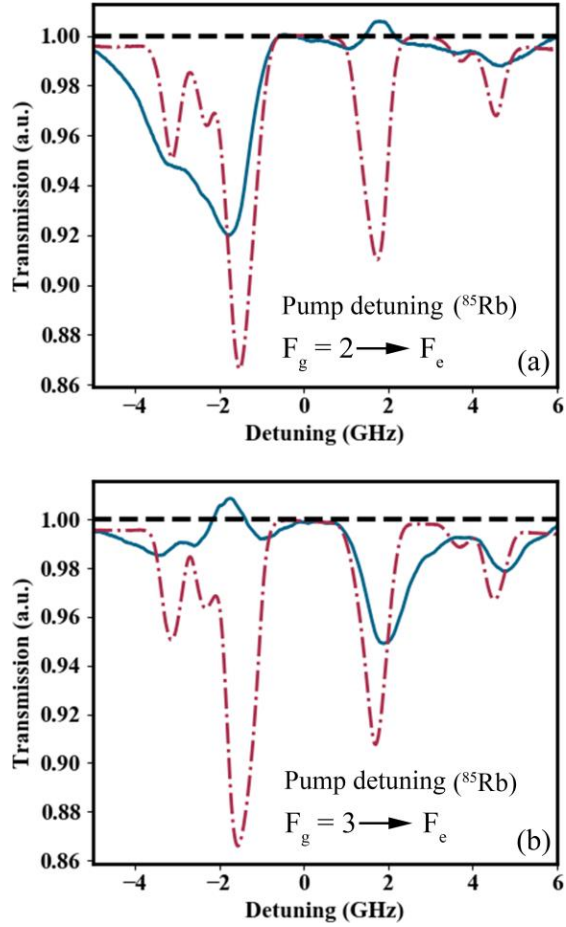

FIG. 6. D1 line transmission for natural rubidium in micrometer vapor cell at temperature of 200°C where the pump beam is fixed at $F_g = 3 \to F_e$ transition (up) and $F_g = 2 \to F_e$ transition of 85Rb. The pump beam (780 nm) intensity ~2 W and the probe intensity is around 150 µW (795 nm) Red dotted line represents the transmission of reference cell at room temperature.

Using the above-mentioned setup, we have measured the transmission spectrum of light through the miniaturized vapor cell, as shown in Fig. 6. The red dotted lines in fig. 6 represent the transmission spectra of the $D_1$ transition of natural Rb vapor measured in a 7.5 cm long reference cell at room temperature (25°C), while the blue lines represent the transmission spectra as obtained from our 30-micron thick cell at a temperature of 200°C. In the top panel (Fig. 6a) the pump was parked at the $F_g = 2 \to F_e$ transition of the 85Rb, whereas at the bottom panel (Fig. 6b), the pump was parked on the $F_g = 3 \to F_e$ of the 85Rb transition. Due to the use of the thin cell resulting in a low optical density, the overall contrast is approximately 5% even at these elevated temperatures. The line width of transmission spectra obtained from the 30-micron cell is broader than the transmission measured from the reference cell, mainly because of power broadening and because of increased Doppler broadening (due to the elevated temperatures) and pressure broadening (due to the presence of the buffer gas).

In both cases (Fig. 6a and Fig. 6b), we observe gain in the transmission spectra. The observed gain from our 30 micron long cell is approximately 1%, validating our assumption of inversionless gain. As predicted, the spectral lines in our cells are broader than the spectral lines in the reference cell. This is attributed to the combined effect of pressure broadening, power broadening and enhanced Doppler broadening.

In Fig. 7 we repeat the same experiments, however this time the pump beam is set to the $^{87}$Rb resonance transitions. In order to reach significant contrast in the $^{87}$Rb (only ~27% in natural rubidium), we need to increase the optical density. To do so, the temperature was increased to 250°C. Indeed, when the pump was aligned to the $F_g = 1 \rightarrow F_e$ (Fig. 7a) the observed absorption contrast is ~ 12% and the gain is ~ 2%. Furthermore, for the stronger transition $F_g = 2 \rightarrow F_e$ (Fig. 7b) the observed gain is approaching 4%.

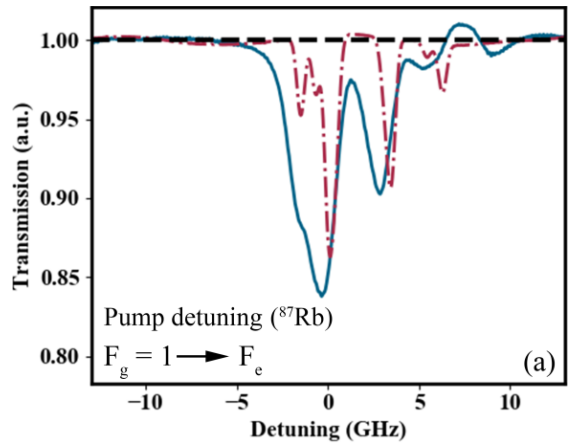

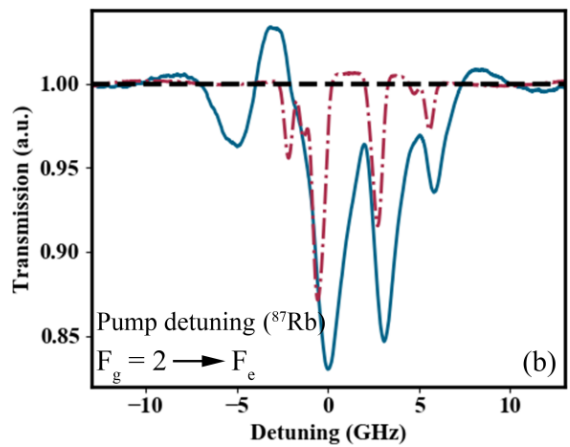

FIG. 7. D1 line transmission for natural rubidium in micrometer vapor cell at temperature of 250°C where the pump beam is fixed at $F_g = 1 \to F_e$ transition (a) and $F_g = 2 \to F_e$ transition of 87Rb. The pump beam (780 nm) intensity ~2 W and the probe intensity is around 150 μW (795 nm). Red dotted line represents the transmission of reference cell at room temperature.

In order to verify that the gain results from coherence rather than from population inversion, we repeated the results of Fig. 6a, however this time we increased the laser linewidth by modulating the current of diode laser with a noisy signal. By doing so, the linewidth is increased allowing us to destroy the coherence, while maintaining the population distribution. In Fig. 8 we present the transmitted probe signal with (blue curve) and without (green curve) laser modulation. As can be seen, only the green curve shows gain, whereas the gain disappears for the blue curve.

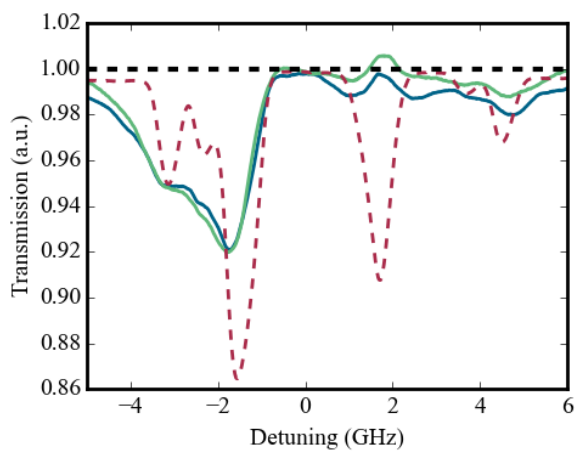

FIG. 8. D1 line transmission of natural rubidium micrometer vapor cell without modulation (green line) and with modulation (blue line). The parameters are temperature of 200°C, pump beam (780 nm) intensity ~2 W and probe intensity is around 150 μW (795 nm). The pump is aligned to the $F_g = 2 \to F_e$ transition of $^{85}$Rb. Red dotted line represents the transmission of the probe through a reference cell at room temperature.

## Conclusion

We have demonstrated theoretically and experimentally the existence of gain without population inversion in miniaturized vapor cells of Rubidium. The effect is a direct result of coherence between excited states. To achieve the condition of gain without population inversion we use buffer gas and miniaturize vapor cell. The use of buffer gas increases the population transition rate from level 4 to level 3, whereas the miniaturized cell results in enhancement of the collision rate, destroying the optical pumping and allows constant transition from the ground state to level 4 by the pump. Our results may pave the way to the demonstration of laser without inversion in miniaturized vapor cells.

## Acknowledgment

E.T acknowledges financial support from the center for nanoscience and nanotechnology of the Hebrew University. The research was supported by a grant from the Israeli Science Foundation (ISF).

## Appendix A Liouville master equation:

The equations for the atomic density matrix elements take the form

$$\partial_t \rho_{11} = \frac{1}{2}\left(2W_{12}(-\rho_{11}+\rho_{22}) + \gamma_3\rho_{33} + \gamma_4\rho_{44} - 2i\left((\rho_{13}-\rho_{31})\Omega_{\text{pr}} + (\rho_{14}-\rho_{41})\Omega_{\text{pu}}\right)\right)$$

$$\partial_t \rho_{12} = -(W_{12}+i\Delta_{\text{HFS}})\rho_{12} + i(\rho_{32}\Omega_{\text{pr}} + \rho_{42}\Omega_{\text{pu}})$$

$$\partial_t \rho_{13} = -\frac{1}{2}(R+W_{12}+\gamma_3+2i\Delta_{\text{pr}})\rho_{13} - i(\rho_{11}-\rho_{33})\Omega_{\text{pr}} + i\rho_{43}\Omega_{\text{pu}}$$

$$\partial_t \rho_{14} = -\frac{1}{2}(R+W_{12}+\gamma_4)\rho_{14} + i(\rho_{34}\Omega_{\text{pr}} + (-\rho_{11}+\rho_{44})\Omega_{\text{pu}})$$

$$\partial_t \rho_{21} = -(W_{12}-i\Delta_{\text{HFS}})\rho_{21} - i(\rho_{23}\Omega_{\text{pr}} + \rho_{24}\Omega_{\text{pu}})$$

$$\partial_t \rho_{22} = \frac{1}{2}(\gamma_3\rho_{33} + \gamma_4\rho_{44})$$

$$\partial_t \rho_{23} = -\frac{1}{2}(R+W_{12}+\gamma_3-2i\Delta_{\text{HFS}}+2i\Delta_{\text{pr}})\rho_{23} - i\rho_{21}\Omega_{\text{pr}}$$

$$\partial_t \rho_{24} = -\frac{1}{2}(R+W_{12}+\gamma_4-2i\Delta_{\text{HFS}})\rho_{24} - i\rho_{21}\Omega_{\text{pu}}$$

$$\partial_t \rho_{31} = -\frac{1}{2}(R+W_{12}+\gamma_3-2i\Delta_{\text{pr}})\rho_{31} + i((\rho_{11}-\rho_{33})\Omega_{\text{pr}} - \rho_{34}\Omega_{\text{pu}})$$

$$\partial_t \rho_{32} = -\frac{1}{2}(R+W_{12}+\gamma_3+2i\Delta_{\text{HFS}}-2i\Delta_{\text{pr}})\rho_{32} + i\rho_{12}\Omega_{\text{pr}}$$

$$\partial_t \rho_{33} = -(R+\gamma_3)\rho_{33} + R\rho_{44} + i(\rho_{13}-\rho_{31})\Omega_{\text{pr}}$$

$$\partial_t \rho_{34} = -\frac{1}{2}(2R+\gamma_3+\gamma_4-2i\Delta_{\text{pr}})\rho_{34} + i(\rho_{14}\Omega_{\text{pr}} - \rho_{31}\Omega_{\text{pu}})$$

$$\partial_t \rho_{41} = -\frac{1}{2}(R+W_{12}+\gamma_4)\rho_{41} - i(\rho_{43}\Omega_{\text{pr}} + (-\rho_{11}+\rho_{44})\Omega_{\text{pu}})$$

$$\partial_t \rho_{42} = -\frac{1}{2}(R+W_{12}+\gamma_4+2i\Delta_{\text{HFS}})\rho_{42} + i\rho_{12}\Omega_{\text{pu}}$$

$$\partial_t \rho_{43} = -\frac{1}{2}(2R+\gamma_3+\gamma_4+2i\Delta_{\text{pr}})\rho_{43} + i(-\rho_{41}\Omega_{\text{pr}} + \rho_{13}\Omega_{\text{pu}})$$

$$\partial_t \rho_{44} = R\rho_{33} - (R+\gamma_4)\rho_{44} + i(\rho_{14}-\rho_{41})\Omega_{\text{pu}}$$

The closeness of the system requires $\rho_{11}+\rho_{22}+\rho_{33}+\rho_{44}=1$. For the convenience of the calculation, $W_{12}$ and R and $\Omega_{\text{pr}}, \Omega_{\text{pu}}$ are chosen to be real.

## Appendix B: collisional transfer between $5^2P_{1/2,3/2}$

The energy transfer between spin orbit split states, with rate coefficient $R_{34}$ and $R_{43}$, via collisions with buffer gas B is the primary process of interest:

(B1) 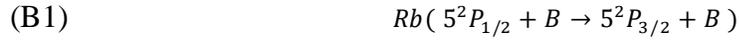

$$Rb(\,5^2P_{1/2} + B \rightarrow 5^2P_{3/2} + B\,)$$

(B2) 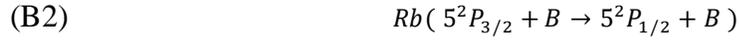

$$Rb(\,5^2P_{3/2} + B \rightarrow 5^2P_{1/2} + B\,)$$

The coefficient rates equal to $R_{34} = n \cdot \sigma_1 \cdot v_{av}$ and $R_{43} = n \cdot \sigma_2 \cdot v_{av}$ where n is the atomic density and $v_{av}$ is the average relative thermal velocity $v_{av} = \sqrt{8kT/\pi\mu}$ where k is the Boltzmann constant and T is the temperature in Kelvin and $\mu$ is the reduced mass given by the masses $m_{Rb}$ of the rubidium atom and the $m_{H_2}$ of the $H_2$ buffer gas molecule is

(B3)
$$\frac{1}{\mu} = \frac{1}{m_{Rb}} + \frac{1}{m_{H_2}}$$

The collisional cross section is given by Table B1

| Collision partner | $\sigma_1$ ($10^{-16}\,cm^2$) | $\sigma_1$ ($10^{-16}\,cm^2$) | T (K) | Ref |
|---|---|---|---|---|
|  | $10 \pm 1.2$ | $13.9 \pm 1.7$ | 330 | [27] |
| $H_2$ | 11 | 15 | 340 | [27] |
|  | >50 | >30 | 1720 | [27] |

## Appendix C - Wall relaxation calculation:

Ground state population relaxation time $T_1$ in a high vacuum cell, in which interatomic collisions are irrelevant, is presented in ref [28], where this relaxation time is identified with the mean time of flight of the atoms between two collisions with the wall. Following these calculations, for a rectangular cell with length l and width h and thickness w, the ground state relaxation rate is expressed as

$$W_{12} = 2\pi \cdot \frac{1}{T_1} = 2\pi \cdot \frac{\bar{v}S}{4V} = 2\pi \cdot \frac{\bar{v} \cdot 2(LW + LH + WH)}{4 \cdot LWH}$$

Where V and S are the cell volume and surface area respectively. $\bar{v}$ is the average velocity which equal:

$$\bar{v} = \sqrt{\frac{8k_B T}{\pi m}}$$

$k_B$ is the Boltzmann constant, T is the temperature and m is the rubidium mass.